\documentclass[twocolumn,prd,showpacs]{revtex4-1}
\usepackage{amsmath,amssymb,pgf,graphicx,color,url}
\newcommand{\be}{\begin{equation}}
\newcommand{\ee}{\end{equation}}

\begin{document}
\title{Constraints on millicharged particles from Planck}

\author{
A.\,D.\,Dolgov$^{a,b,c,d}$,
S.\,L.\,Dubovsky$^{e}$,
G.\,I.\,Rubtsov$^{f}$, and
I.\,I.\,Tkachev$^{f}$}
\affiliation{
$^a$Laboratory of Cosmology and Elementary Particles, Novosibirsk State University, Pirogov street 2, 630090 Novosibirsk, Russia\\
$^b$Dipartimento di Fisica, Universit`a degli Studi di Ferrara,  Polo Scientifico e Tecnologico - Edificio C, Via Saragat 1, 44122 Ferrara, Italy\\
$^c$Istituto Nazionale di Fisica Nucleare, Sezione di Ferrara,
Polo Scientifico e Tecnologico - Edificio C, Via Saragat 1, 44122 Ferrara, Italy
\\
$^d$Institute of Theoretical and Experimental Physics,
Bolshaya Cheremushkinskaya ul. 25, 113259 Moscow, Russia\\
$^e$ Center for Cosmology and Particle Physics, Department of Physics, New York University, New York, NY, 10003, USA\\
$^f$Institute  for Nuclear Research of the Russian Academy of Sciences,
Moscow 117312, Russia
}

\begin{abstract}
We revisit cosmic microwave background (CMB) constraints on the
abundance of millicharged particles based on the Planck data. The
stringent limit $\Omega_{mcp}h^2 < 0.001$\,(95\%\,CL) may be set
using the CMB data alone if millicharged particles participate in the
acoustic oscillations of baryon-photon plasma at the recombination
epoch. The latter condition is valid for a wide region of charges and
masses of the particles. Adding the millicharged component to
$\Lambda$CDM shifts prefered scalar spectral index of  primordial perturbations to somewhat larger values as compared to minimal model, even approaching Harrison-Zeldovich spectrum under some assumptions.
\end{abstract}
\maketitle

\section{Introduction}

The nature of dark matter is unknown. Suggested particle candidates range
from massive gravitons, through axions, sterile neutrino, wimps, mirror matter,   wimpzillas, and up to primordial black holes (for a recent reviews see e.g. \cite{Bertone:2010zza}), with the common requirement that they should interact rather weakly. In particular, the dark matter constituents cannot carry electric charge comparable to that of electrons, unless
they are extremely heavy. This does not preclude though the
intriguing possibility of electrically millcharged particles and
corresponding models were discussed extensively in the past.

Millicharged particles have electric charge $e' = \epsilon e $, where
$e$ is the electron charge and $\epsilon \ll 1$. They can be either
bosons or fermions. Nothing prevents their ad hoc introduction into
the theory, nevertheless, they appear naturally in a wide class of
models. Consider e.g. a model with a hidden gauge sector. It may contain a 
$U(1)$ gauge group. The corresponding  `dark photon' field $A'_\mu$ may have a kinetic
mixing, $\frac{\epsilon}{2} F'_{\mu\nu} F^{\mu\nu}$, to `our' photon
$A_\mu$ \cite{Holdom}. This mixing makes shadow charge carriers to be
millicharged with respect to our world. Moreover, these charge carriers
can be `mirror' replicas of our electrons and protons. This opens up a
possibility of dark matter being the `mirror world' which is a shadow
sector with particles having strong and electroweak interactions
similar to the ordinary particles. The mirror world may have different
weak and strong scales \cite{BDM}, or microphysics in it can be
exactly identical to the Standard Model, see a review \cite{Alice} and
references therein.  The existence of models with natural appearance
of millicharged particles grants them close attention.

Bounds on parameters of millicharged particles are derived from direct
laboratory tests as well as from cosmological and astrophysical
observations. These bounds are strongly sensitive to the mass range of
particles involved. In what follows we will call millicharged
particles as $X$-particles.

If $X$-particles are lighter than an electron, $m_X < m_e$, the best
particle physics limit comes from the data on invisible decay mode of
ortho-positronium into $X\bar X$, according to which $\epsilon < 3.4\cdot
10^{-5}$ \cite{positronium}. For very light particles, $m_X < 1$ keV,
the reactor experiments give stronger bound $\epsilon <
10^{-5}$~\cite{Gninenko:2006fi}. 
For $m_X > m_e$ the direct bounds were obtained in Ref.~\cite{Prinz}. For $m_X = 1$ MeV they yield $\epsilon < 4.1 \times 10^{-5}$ while the bounds become weaker for heavier
$X$-particles, up to $\epsilon < 5.8 \times 10^{-4}$ for $m_X=100$
MeV. For $X$-particles heavier than 100 MeV electric charges
$e' \sim 10^{-2} e$ are allowed, while for $m_X > 1$ GeV they can be as
large as $\epsilon=0.1$.

Stellar evolution provides very restrictive limits on $\epsilon$, see
Refs.~\cite{raffelt,milli-limits}, but they are unapplicable if $m_X > 10$
keV. On the other hand millicharged particles may survive late time annihilation by forming  bound states with protons and $\alpha$-particles. Therefore terrestrial 
searches result in the constraint $\epsilon < 0.01$ for
$m_X>1$~GeV~\cite{Langacker:2011db}. 

The presence of millicharged particles  in the early Universe during the epoch of big bang nucleosynthesis (BBN) influences the standard cosmological picture in several respects. In particular, both the expansion rate of the Universe and baryon-to-photon ratio can be significantly altered during BBN and this is dangerous
\cite{milli-limits}. The impact of such particles on BBN is discussed
in Ref~\cite{BDT-BBN}, with the conclusion that the  BBN upper bound on the charge of light millicharged particles can be avoided assuming non-zero lepton asymmetry.

Sufficient abundance of millicharged particles in the "late" Universe
(not necessarily making the whole of dark matter) may lead to a
profound cosmological consequences also. These implications may be even
`useful'. E.g., in the Ref.~\cite{Berezhiani:2013dea} it has been
suggested that the difference in electromagnetic drag forces on
electrons and protons from the millicharged matter during the process
of galaxies formation may explain the long standing puzzle of the
origin of seeds for galactic and cluster magnetic fields.

On the other hand, if millicharged particles  have  sufficiently strong coupling
to baryons and participate in the acoustic oscillations of
baryon-photon plasma at the recombination epoch \cite{Dubovsky:2001tr}
then the power spectrum of Cosmic Microwave Background (CMB) radiation 
anisotropies is affected  in several ways. For $l \lesssim 1000$ the contribution of millicharged particles is degenerate with that of baryonic matter. Using this fact, and employing nucleosynthesis data on baryon abundance together with WMAP CMB spectrum,  a strong limit on cosmological abundance of millicharged particles was obtained in  Ref.~\cite{Dubovsky:2003yn}. At $l \gtrsim 1000$  millichared particles lead to direct additional suppression of the spectrum. Recent Planck satellite
data~\cite{Planck_overview} are  precise at high multipoles up to $l \sim 2500$. Therefore, Planck spectra can be used to look for this effect of additional suppresion. The aim of the present paper is to consider corresponding CMB anisotropy limits on millicharged particles in view of the Planck data.

\section{CMB constraints on millicharged matter abundance}

Millicharged particles scatter off electrons and protons at the
recombination epoch. It was shown that if the velocity transfer rate
of this process exceeds the expansion rate of the Universe, the
millicharged particles behave similar to baryons until
recombination~\cite{Dubovsky:2001tr}.  The tight coupling condition is
given by~\cite{Dubovsky:2003yn}:
\begin{equation}
\label{condition}
\Gamma_{mcp} (\Omega_b +
 \Omega_{mcp}) H^{-1} \gtrsim 250\;,
\end{equation}
where $\Omega_b$ and $\Omega_{mcp}$ are baryon and X-particles
abundances correspondingly, $H$ is the Hubble parameter and
$\Gamma_{mcp}$ is a velocity tranfer rate at recombination. The latter
is given by the following equation:
\begin{equation}
\Gamma_{mcp} =   \frac{4\sqrt{2\pi} 
\alpha^2 \epsilon^2 \rho_{crit}}{3 m_X m_p T^{3/2}} |\ln \theta_D| 
(\sqrt{\mu_{X,e}} + \sqrt{\mu_{X,p}})\;,
\label{R1}
\end{equation}
where $\mu_{X,e(p)}$ is the reduced mass of a millicharged particle
and electron (proton), $\alpha$ is the fine structure constant, $T$ is
a temperature, $\rho_{crit}$ is the critical density at recombination
and $\theta_D =\sqrt{2\pi\alpha n_e/T^2m_e}$ is the Debye angle. The
tight coupling condition given by Eq.~(\ref{condition}) is valid in the
broad region of the parameter space~
\begin{equation}
\epsilon^2 \gtrsim 5\cdot 10^{-11}\,\mbox{GeV}^{-1/2}
\frac{m_X}{\sqrt{\mu_{X,e}}+\sqrt{\mu_{X,p}}}.
\label{R2}
\end{equation}

Unlike baryons, millicharged particles do not contribute to the plasma
opacity at the recombination since the Compton process is suppressed by
the fourth power of the charge. Therefore the photon mean free path before recombination
is longer if a fraction of baryons is replaced by millicharged particles and  high multipoles of the CMB spectrum get suppressed. Consequently, as proposed in~\cite{Dubovsky:2003yn}, an accurate determination
of high CMB miltipoles should result in strong limits on
the millicharged particles. The Planck collaboration has recently
published\,\cite{Planck_overview} high-quality measurements of the CMB spectrum up to
$l \sim 2500$.  In this note we use the
Planck data on CMB to set a limit on the abundance of millicharged
particles. We assume that the millicharged particles are stable and
that the tight coupling condition holds.

We numerically solve the linearized kinetic equations for the
primordial plasma in synchronous gauge with
CAMB~\cite{CAMB}. The millicharged component and the
corresponding equations are added to CAMB similarly to the
modification of CMBFAST~\cite{Seljak:1996is} performed in
Ref.~\cite{Dubovsky:2003yn}.

We extend the spatially-flat six-parameter $\Lambda$CDM
cosmology~\cite{Planck_params} with one additional parameter: the
present abundance of millicharged particles $\Omega_{mcp} h^2$. { We start with 
assumption that neutrinos are  effectively massless, i.e. the mass of
one neutrino is  $m_\nu=0.06$\,eV, and the effective number of neutrinos is $n_\nu = 3.046$. The parameter space is explored using the Markov Chain Monte-Carlo technique with the COSMOMC (March
2013) package~\cite{COSMOMC}. The Planck likelihood code
\url{plc-1.0} is used~\cite{Planck_powerspec} in conjunction with the
fast-slow sampling method~\cite{COSMOMC_fastslow}. The latter
technique performs an efficient sampling over the subclass of ``fast''
parameters which do not require to recalculate the CMB transfer functions
(e.g., the overall normalization of the spectrum and the Planck calibration
parameters).

\begin{figure*}
\includegraphics[angle=270,width=0.48\textwidth]{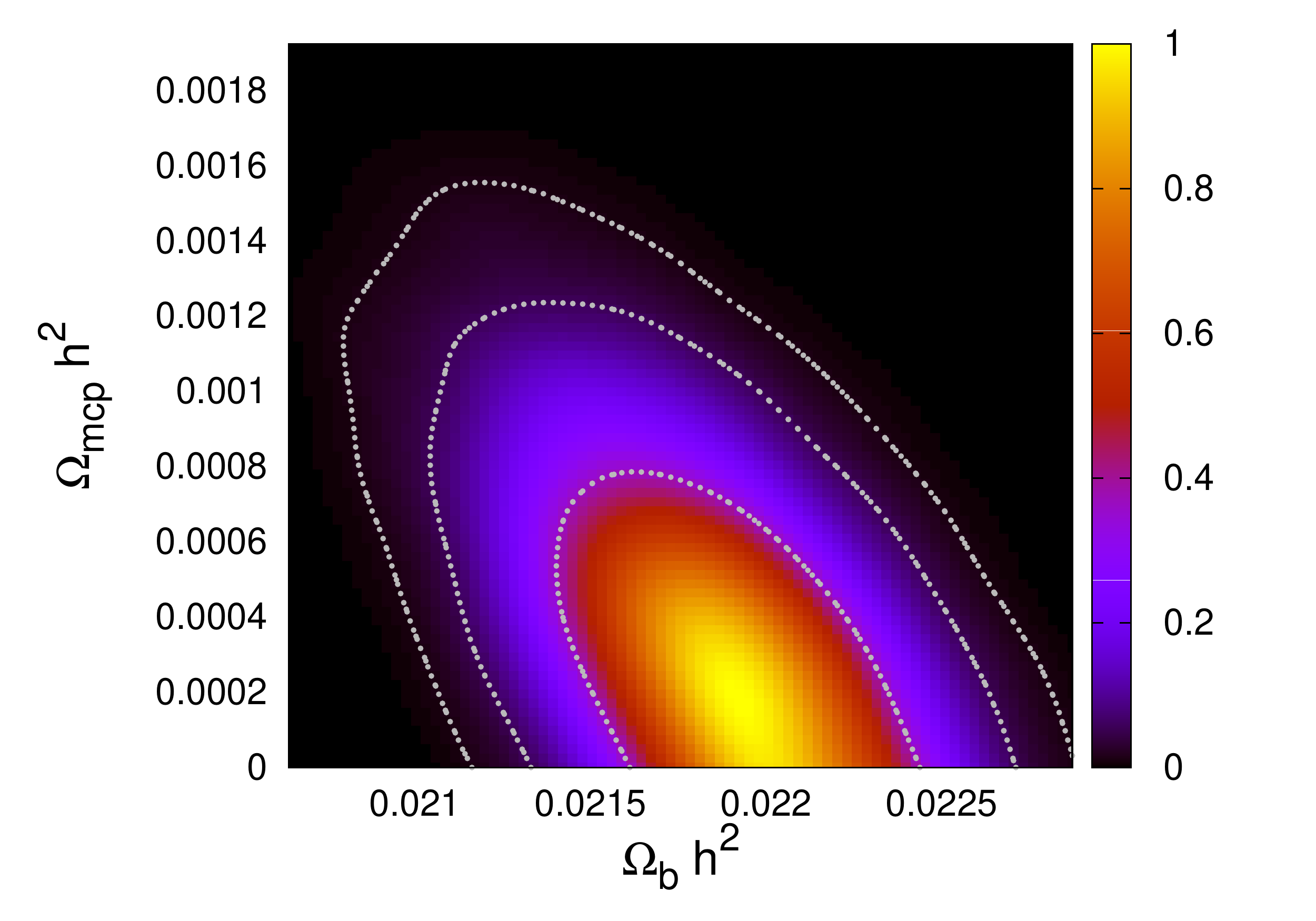}
\includegraphics[angle=270,width=0.48\textwidth]{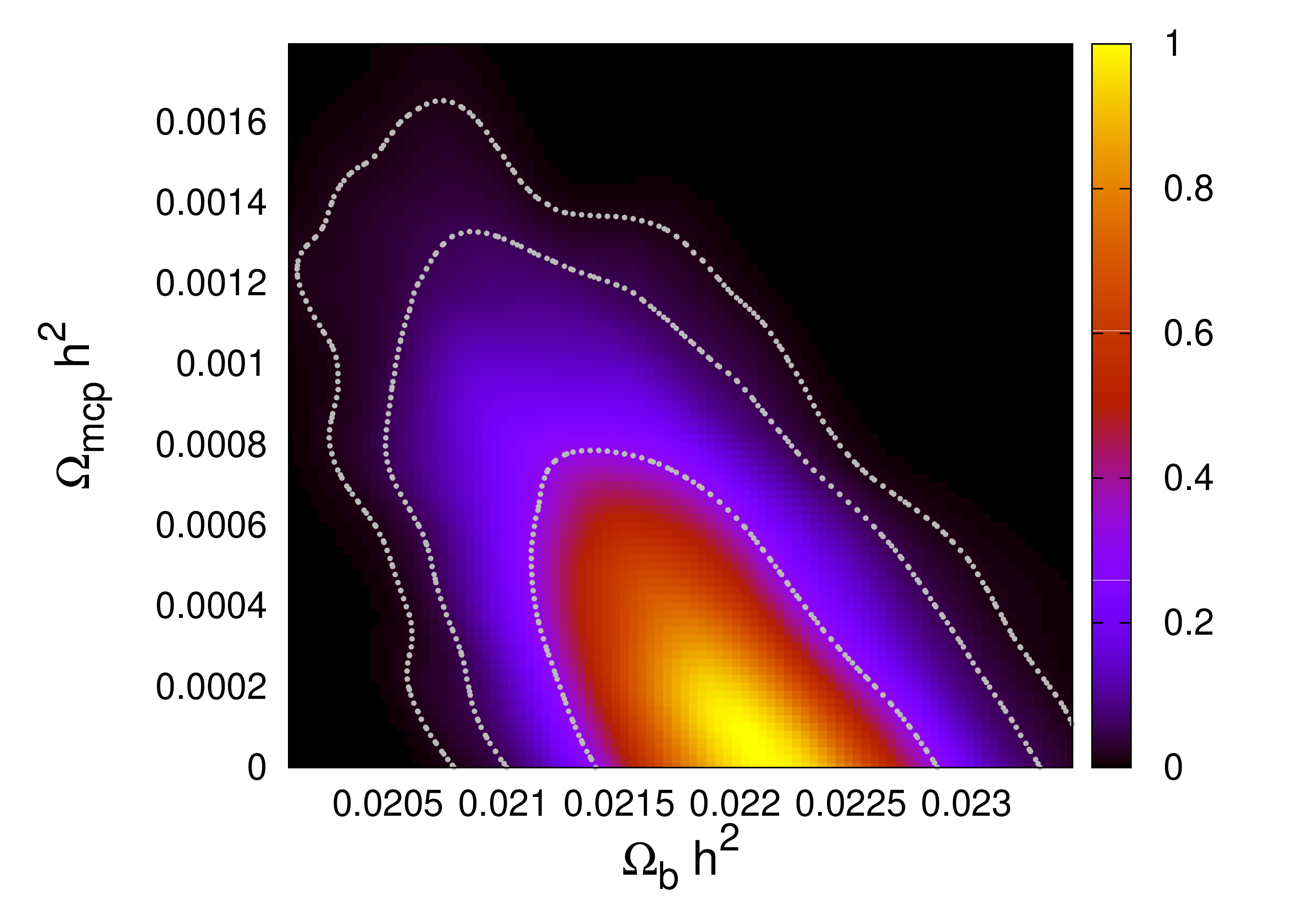}
\caption{
The marginalized likelihood in the $\Omega_{mcp} h^2$ - $\Omega_b h^2$ plane. The countours constrain 68\%, 95\% and 99\% regions. Left panel corresponds to 3 effectively massless neutrino species, see text for exact values of these parameters. In the right panel the constraints on the number and mass of neutrinos are relaxed. 
}
\label{mcp_likeli}
\end{figure*}

\begin{figure*}
\includegraphics[angle=270,width=0.48\textwidth]{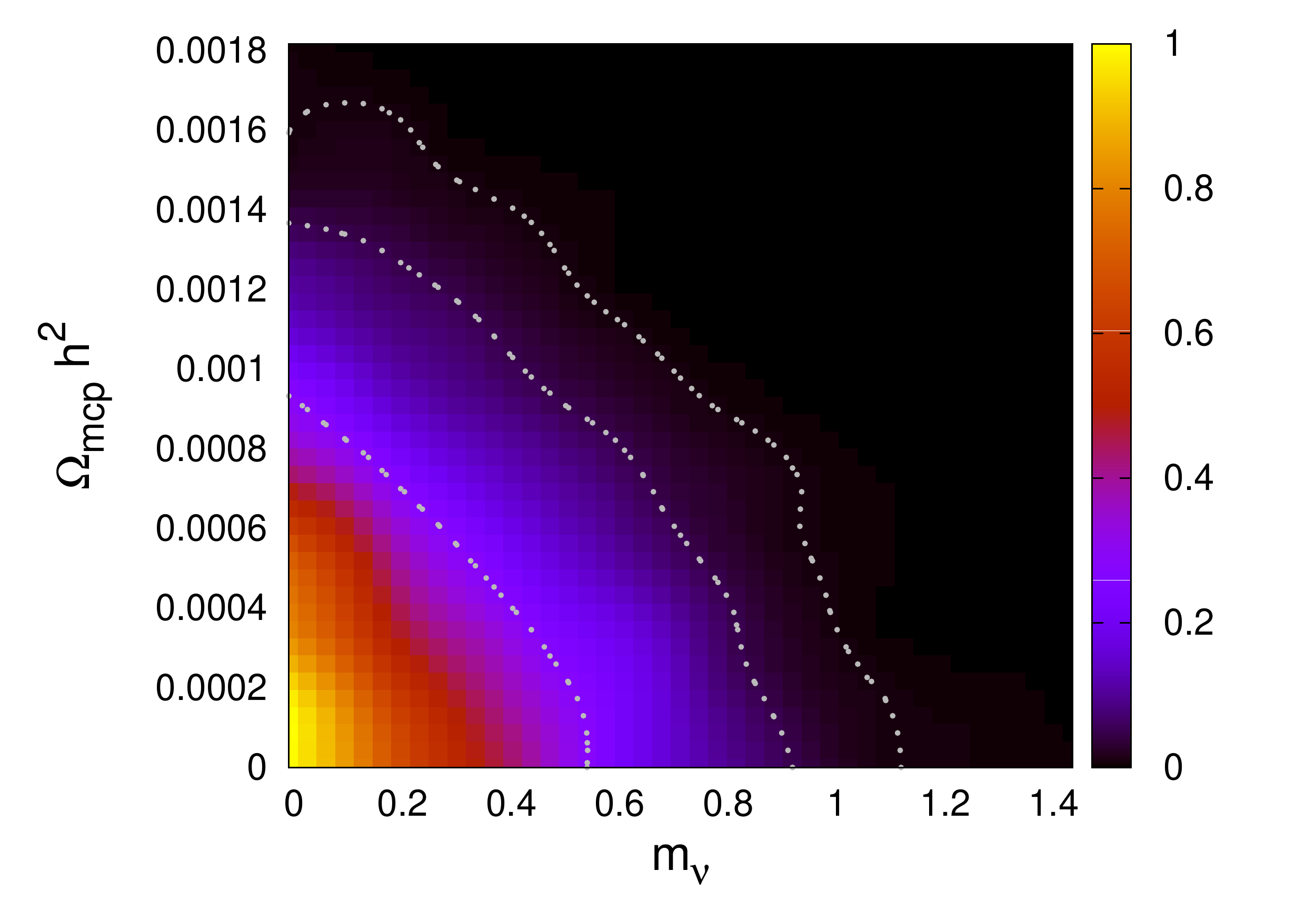}
\includegraphics[angle=270,width=0.48\textwidth]{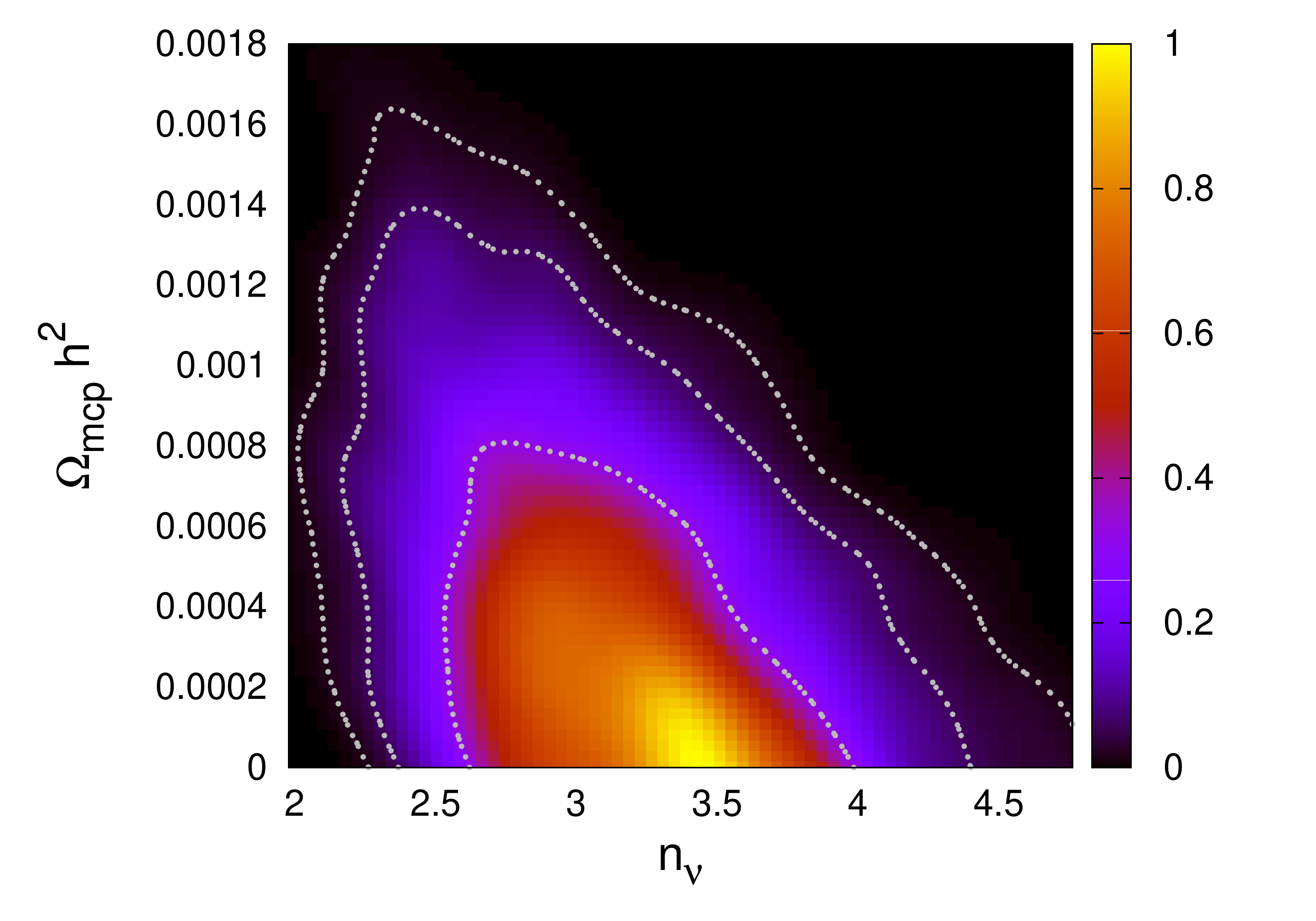}
\caption{\label{numcp_likeli}
The marginalized likelihood in the 2D planes: $\Omega_{mcp} h^2$ versus $m_\nu$,
and $n_\nu$. The countours constrain 68\%, 95\% and 99\% regions.
}
\end{figure*}

We generated 10 Markov chains with a total of $3\times 10^5$ samples
using a covariance matrix supplied with COSMOMC. The marginalized
likelihood in the $\Omega_{mcp} h^2$ - $\Omega_b h^2$ plane of parameters is shown in
Figure~\ref{mcp_likeli}. Left panel corresponds to the above quoted fixed values 
for the number and mass of neutrinos.

We arrive at the following upper limit

\be \label{lim} \Omega_{mcp}h^2 < 0.001\,(95\%\,\mbox{CL}) \ee

An approximate degeneracy in the parameter space is seen in
Fig.~\ref{mcp_likeli}, namely, larger  $\Omega_{mcp} h^2$ leads to a smaller $\Omega_b h^2$. The degeneracy agrees with that 
predicted in~\cite{Dubovsky:2003yn} based on hypothetical CMB measurements up to $l=1600$. Planck measures CMB 
spectrum up to higher multipoles and
hence the limit is stronger than expected. Note that Planck provides
much more precise CMB data than WMAP and, as expected, the Big Bang
nucleosynthesis data~\cite{TODO:latest_bbn_omega_b} don't help to improve
the constraints.

To test the robustness of the result we relaxed the number of
effective neutrinos and the mass of one neutrino. We generated ten new
Markov chains with a total of $1\times 10^5$ samples.  The
marginalized likelihood in $\Omega_{mcp} h^2$ - $\Omega_b h^2$ plane of parameters is shown in Fig.~\ref{mcp_likeli}, right panel. Likelihoods for $\Omega_{mcp} h^2$ versus $m_\nu$ and $n_\nu$ are shown in Fig.~\ref{numcp_likeli}, left and right panels respectively. Overall limit on the millicharged particles abundance  after relaxing the assumptions about neutrino is still given by Eq.~(\ref{lim}) since the  increase in $m_\nu$ or $n_\nu$ leads to a smaller allowed fraction of millicharges.


\section{Millicharged matter and constraints on Inflation}

\begin{figure*}
\includegraphics[angle=270,width=0.48\textwidth]{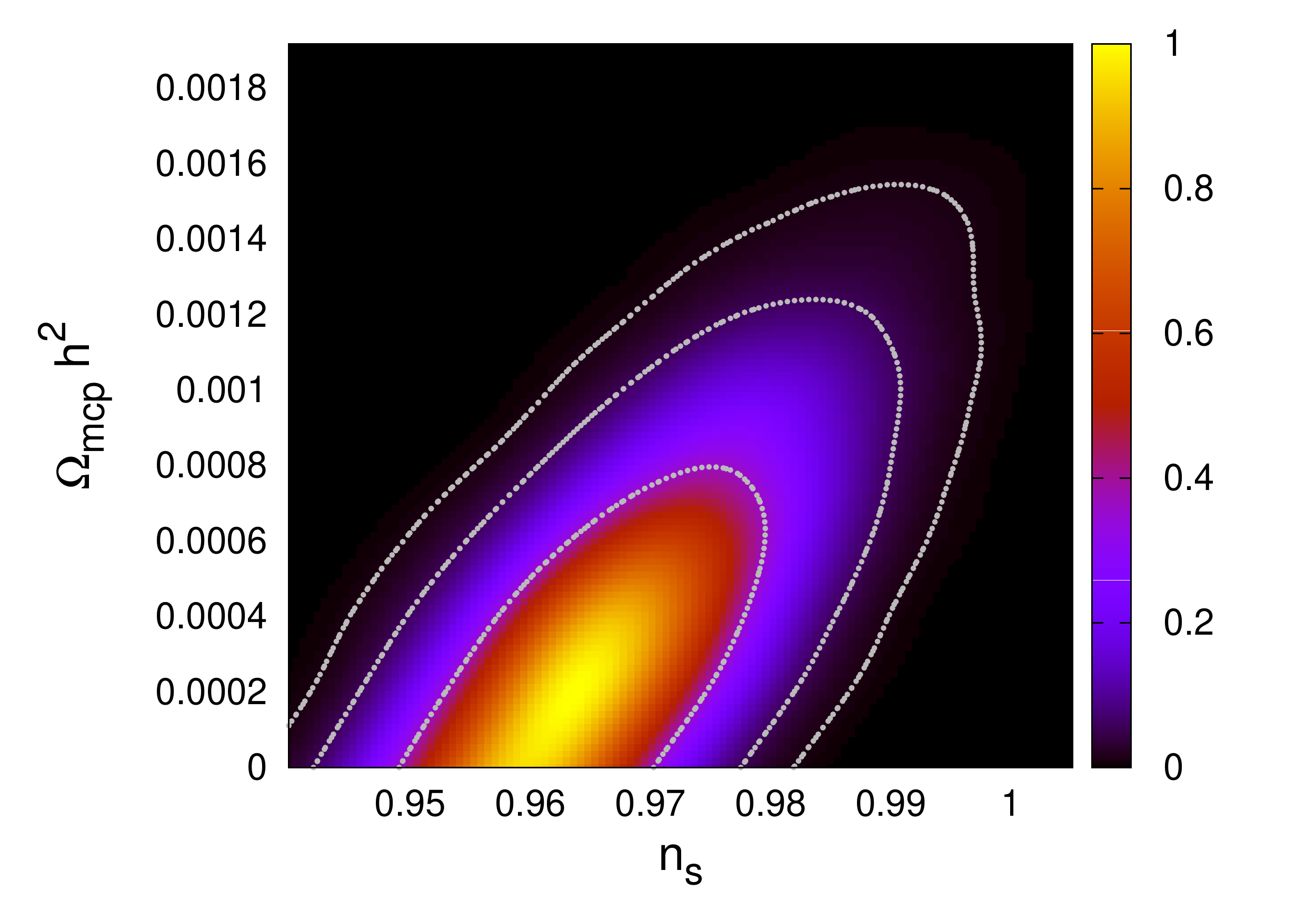}
\includegraphics[angle=270,width=0.48\textwidth]{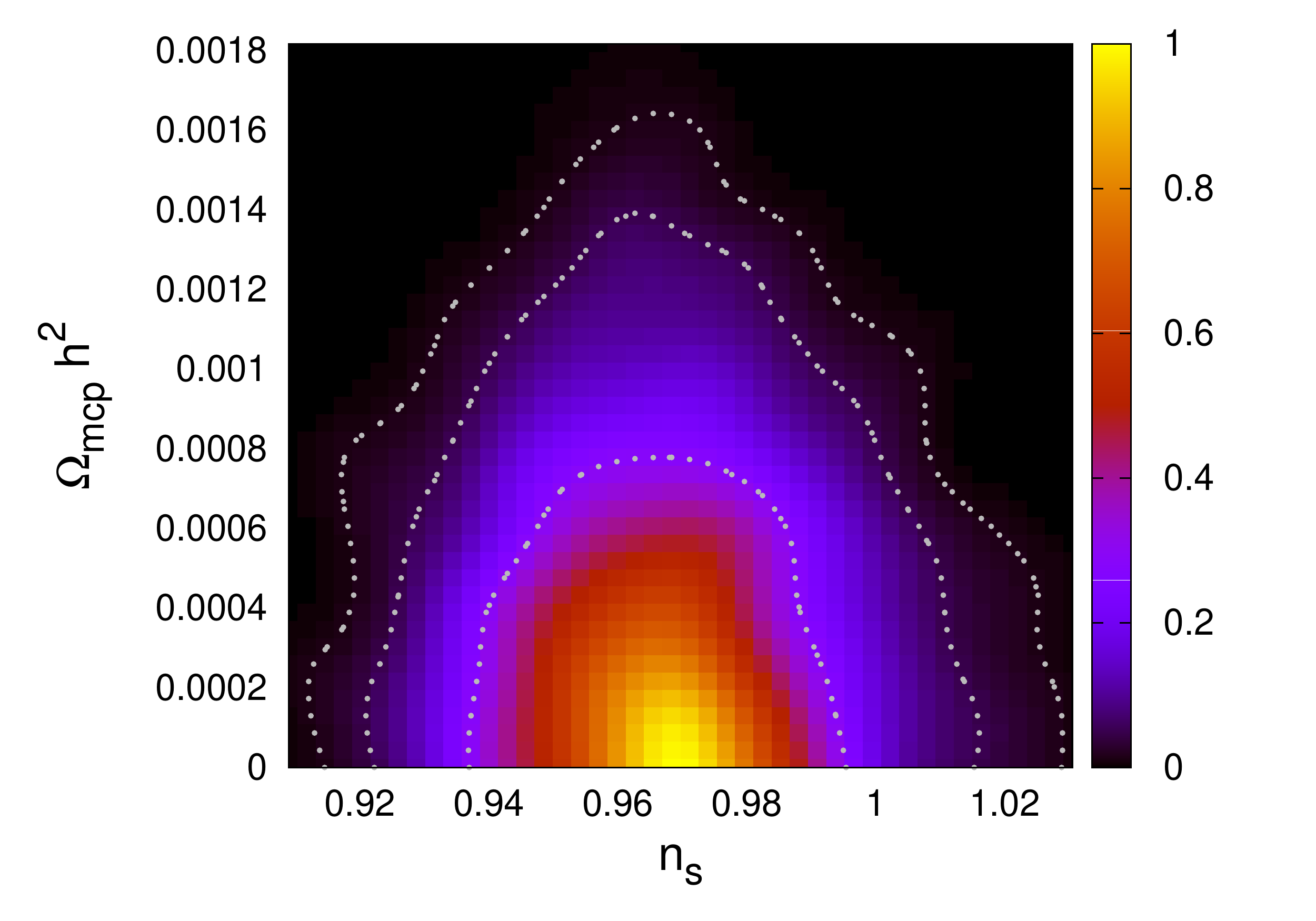}
\caption{
Same as in Fig.~\ref{mcp_likeli} but in the $\Omega_{mcp} h^2$ - $n_s$ plane. 
}
\label{mcp_n_s_likeli}
\end{figure*}

Precision results of the Planck mission lead to the conclusion that the  power-law index of the spectrum  of scalar primordial perturbations  in the standard cosmological  (aka $\Lambda$CDM) and Standard particle physics models is restricted to the value $n_s = 0.9603 \pm 0.0073$~\cite{Planck_params},   which is about $6\sigma$ away from scale 
invariant Harrison-Zeldovich  spectrum. At the face value this would mean also that in this simplest setup many models of inflation are disfavored \cite{Ade:2013uln} with respect to  e.g. $R^2$ ~\cite{Starobinsky:1980te} or Higgs~\cite{Bezrukov:2007ep} models of inflation. This result is of great importance for cosmology. Therefore, it should be inspected how this conclusion holds with respect to the possible extensions of the Standard Model of particle physics.

In the presence of millicharged matter the higher multipoles of the CMB spectrum are
additionally suppressed, which clearly should change the derived value of $n_s$. To quantify this we plot  in Fig.~\ref{mcp_n_s_likeli} the marginalized likelihood of parameters in the $\Omega_{mcp} h^2$ - $n_s$ plane. Left panel in this Figure corresponds to the case of `3 massless' neutrinos. We see that, indeed, with the increase of $\Omega_{mcp}$ the value of $n_s$ increases, pushing it out from $R^2$ inflation into the range predicted by other models, such as the spontaneously broken SUSY model of Ref.~\cite{Dvali:1994ms}.

Relaxing the constraints on the number of neutrinos widens the allowed range for $n_s$ even further, and now even the Harrison-Zeldovich  spectrum is not ruled out, see Fig.~\ref{mcp_n_s_likeli}, right panel. However, the preferred value of $n_s$ does
not depend upon $\Omega_{mcp} h^2$.

\section{Conclusions}

We find that the cosmological abundance of millicharged particles is strongly constrained by the Planck data, $\Omega_{mcp} h^2 < 0.001$. The mechanism for generation of galactic magnetic  seed filelds, proposed in~\cite{Berezhiani:2013dea}, is not ruled out by this limit. The precise value of the limit depends upon the particle physics model assumed. E.g. the maximally allowed  value for the  abundance, at the level of $\Omega_{mcp} h^2 \approx 0.001$, is incompatible with the assumption of 4 light neutrinos, while it can be achieved if only 3 `massless' neutrinos are present.

In addition, the 
abundance of millicharges is partially degenerate with the power-law index of primordial scalar perturbations, $n_s$. 
Existence of millicharge particles reopens the possibility for some inflationary scenario,   
such as the spontaneously broken SUSY model of inflations, which would be excluded under minimal assumptions.

\paragraph*{Acknowledgements} AD, GR and  IT acknowledge  support of the Russian Federation Government Grant No. 11.G34.31.0047.
The work is supported in part by the RFBR grants 12-02-00653, 12-02-31776,
12-02-91323 and 13-02-01293, by the Dynasty foundation\,(GR), by the
grants of the President of the Russian Federation NS-5590.2012.2, MK-1170.2013.2 (GR). 
The work of SD is partially supported by the NSF grants PHY-1068438 and PHY-1316452.
The numerical part of the work has been done at the cluster of the Theoretical Division of INR RAS.


%
%

\begin{thebibliography}{99}

\bibitem{Bertone:2010zza} 
G.~Bertone (ed.), {\it Particle Dark Matter: Observations, Models and Searches},
  (Cambridge Univ. Press, 2010).

\bibitem{Holdom} 
B. Holdom, Phys. Lett. B {\bf 166}, 196 (1986). 

\bibitem{BDM} 
  Z.~G.~Berezhiani, A.~D.~Dolgov and R.~N.~Mohapatra,
  Phys.\ Lett.\ B {\bf 375}, 26 (1996)
  [hep-ph/9511221].

\bibitem{Alice} 
  Z.~Berezhiani,
  Int.\ J.\ Mod.\ Phys.\ A {\bf 19}, 3775 (2004).
  
\bibitem{positronium}
A. Badertscher et al, 
Phys. Rev. D75 (2007) 032004.

\bibitem{Gninenko:2006fi} 
  S.~N.~Gninenko, N.~V.~Krasnikov and A.~Rubbia,
  Phys.\ Rev.\ D {\bf 75}, 075014 (2007)
  [hep-ph/0612203].

\bibitem{Prinz} 
A.A. Prinz et al., Phys. Rev. Lett. 81, 1175 (1998); 

\bibitem{raffelt}
G.G. Raffelt, {\it Stars as Laboratories for Fundamental Physics}, the University of Chicago Press, 1996.


\bibitem{milli-limits}
S. Davidson, S. Hannestad, G. Raffelt, JHEP 05 (2000) 003;
A. Melchiorri, A. Polosa, A. Strumia, Phys. Lett. B650 (2007) 416.

\bibitem{Langacker:2011db} 
  P.~Langacker and G.~Steigman,
  Phys.\ Rev.\ D {\bf 84}, 065040 (2011).

\bibitem{BDT-BBN} 
  Z.~Berezhiani, A.~Dolgov and I.~Tkachev,
  JCAP {\bf 1302}, 010 (2013)
  [arXiv:1211.4937 [astro-ph.CO]].
  
\bibitem{Berezhiani:2013dea} 
  Z.~Berezhiani, A.~D.~Dolgov and I.~I.~Tkachev,
  arXiv:1307.6953 [astro-ph.CO].

\bibitem{Dubovsky:2001tr} 
  S.~L.~Dubovsky and D.~S.~Gorbunov,
  Phys.\ Rev.\ D {\bf 64}, 123503 (2001)
  [astro-ph/0103122].

\bibitem{Dubovsky:2003yn} 
  S.~L.~Dubovsky, D.~S.~Gorbunov and G.~I.~Rubtsov,
  JETP Lett.\  {\bf 79}, 1 (2004).

\bibitem{Planck_overview} 
  P.~A.~R.~Ade {\it et al.}  [Planck Collaboration],
  arXiv:1303.5062 [astro-ph.CO].

\bibitem{CAMB} 
  A.~Lewis, A.~Challinor and A.~Lasenby,
  Astrophys.\ J.\  {\bf 538}, 473 (2000);
  \url{http://camb.info}.

\bibitem{Seljak:1996is} 
  U.~Seljak and M.~Zaldarriaga,
  Astrophys.\ J.\  {\bf 469}, 437 (1996).

\bibitem{Planck_params} 
  P.~A.~R.~Ade {\it et al.}  [Planck Collaboration],
  arXiv:1303.5076 [astro-ph.CO].

\bibitem{COSMOMC} 
  A.~Lewis and S.~Bridle,
  Phys.\ Rev.\ D {\bf 66}, 103511 (2002).
\url{http://cosmologist.info/cosmomc/}.

\bibitem{Planck_powerspec} 
  P.~A.~R.~Ade {\it et al.}  [Planck Collaboration],
  arXiv:1303.5075 [astro-ph.CO].


\bibitem{COSMOMC_fastslow} 
  A.~Lewis,
  Phys.\ Rev.\ D {\bf 87}, 103529 (2013).

\bibitem{TODO:latest_bbn_omega_b} 
A. Coc, J.-P. Uzan, E. Vangioni,  arXiv:1307.6955.


\bibitem{Ade:2013uln} 
  P.~A.~R.~Ade {\it et al.}  [Planck Collaboration],
  arXiv:1303.5082 [astro-ph.CO].
  
\bibitem{Starobinsky:1980te} 
  A.~A.~Starobinsky,
  Phys.\ Lett.\ B {\bf 91}, 99 (1980).
  
\bibitem{Bezrukov:2007ep} 
  F.~L.~Bezrukov and M.~Shaposhnikov,
  Phys.\ Lett.\ B {\bf 659}, 703 (2008)
  [arXiv:0710.3755 [hep-th]].
  
\bibitem{Dvali:1994ms} 
  G.~R.~Dvali, Q.~Shafi and R.~K.~Schaefer,
  Phys.\ Rev.\ Lett.\  {\bf 73}, 1886 (1994)
  [hep-ph/9406319].

\end{thebibliography}
\end{document}